\documentclass[subeqn,12pt, a4paper]{article} 
\usepackage{amssymb,amsmath,amsfonts,amsthm, amscd, mathrsfs,helvet}
\usepackage{bm, stmaryrd}
\usepackage{graphicx,verbatim}
\usepackage[usenames, dvipsnames]{color}
\usepackage{psfrag}
\usepackage[all]{xy}
\usepackage{tikz}
\usetikzlibrary{arrows,matrix,decorations.markings,shapes.geometric}
\usepackage[normalem]{ulem}
\usepackage{xspace}
\usepackage{srcltx}
\definecolor{lightred}{RGB}{255,127,127}
\definecolor{lightgreen}{RGB}{127,255,127}
\definecolor{lightblue}{RGB}{127,127,255}
\definecolor{linkcolor}{rgb}{0,0,0.6}
\usepackage[nosort]{cite}
\usepackage{pdfpages}

\theoremstyle{plain}

\newtheorem*{theorem*}{Theorem}

\newtheorem*{proposition*}{Proposition}

\newcommand{\tensor}[1]{{\bf \underline{#1}}}

\definecolor{brightBlue}{rgb}{0,0,1}

\definecolor{Violet}{rgb}{0.47,0,1}

\DeclareMathOperator{\tr}{Tr}

\DeclareMathOperator{\Res}{Res}

\def\ha{\mbox{\small $\frac{1}{2}$}}
\def\qa{\mbox{\small $\frac{1}{4}$}}

\def\CC{\mathbb{C}}

\def\P{\mathcal{P}}
\def\R{\mathcal{R}}
\def\L{\mathcal{L}}

\def\1{\tensor{1}}
\def\2{\tensor{2}}
\def\3{\tensor{3}}
\def\4{\tensor{4}}

\def\beq{\begin{equation}}
\def\eeq{\end{equation}}
\def\beqz{\begin{equation*}}
\def\eeqz{\end{equation*}}
\def\bea{\begin{eqnarray}}
\def\eea{\end{eqnarray}}

\def\et{\qquad\mbox{and}\qquad}

\newcommand{\bYB}{\text{bYB}}

\newcommand{\noi}{\noindent}
\newcommand{\dd}{\text{d}}
\newcommand{\p}{\partial}
\newcommand{\g}{\mathfrak{g}}
\newcommand{\s}{\sigma}
\newcommand{\Id}{\text{Id}}

\newcommand{\Ad}{\text{Ad}}

\newcommand{\Gd}{G_{\text{diag}}}
\newcommand{\Ob}{\mathcal{O}}
\newcommand{\Lc}{\mathcal{L}}
\newcommand{\Rc}{\mathcal{R}}

\newcommand{\etat}{\tilde{\eta}}
\newcommand{\at}{\tilde{a}}
\newcommand{\Xt}{\tilde{X}}

\newcommand{\jt}{\tilde{j}}
\newcommand{\Jt}{\tilde{J}}

\newcommand{\RgXt}{\tilde{R}_{\tilde{g}}\tilde{X}}

\newcommand{\gt}{\tilde{g}}
\newcommand{\Rt}{\tilde{R}}
\newcommand{\zt}{\tilde{z}}
\newcommand{\Jc}{\mathscr{J}}
\newcommand{\Jct}{\widetilde{\mathscr{J}}}

\newcommand{\ti}[1]{_{\bm{\underline{#1}}}}

\numberwithin{equation}{section}

\begin{document}
 
\begin{center}
\vspace*{2em}
{\large\bf On the Hamiltonian integrability of the bi-Yang-Baxter $\sigma$-model}\\
\vspace{1.5em}
F. Delduc$\,{}^1$, S. Lacroix$\,{}^1$, M. Magro$\,{}^1$, B. Vicedo$\,{}^2$

\vspace{1em}
\begingroup\itshape
{\it 1) Laboratoire de Physique, ENS de Lyon
et CNRS UMR 5672, Universit\'e de Lyon,}\\
{\it 46, all\'ee d'Italie, 69364 LYON Cedex 07, France}\\
\vspace{1em}
{\it 2) School of Physics, Astronomy and Mathematics,
University of Hertfordshire,}\\
{\it College Lane,
Hatfield AL10 9AB,
United Kingdom}
\par\endgroup
\vspace{1em}
\begingroup\ttfamily
Francois.Delduc@ens-lyon.fr, Sylvain.Lacroix@ens-lyon.fr, Marc.Magro@ens-lyon.fr, Benoit.Vicedo@gmail.com
\par\endgroup
\vspace{1.5em}
\end{center}

\begin{abstract}
The bi-Yang-Baxter $\sigma$-model is a certain two-parameter deformation of the principal chiral model on a real Lie group $G$ for which the left and right $G$-symmetries of the latter are both replaced by Poisson-Lie symmetries. 
It was introduced by C. {Klim$\check{\text{c}}$\'{\i}k} who also recently showed it admits a Lax pair, thereby proving it is integrable at the Lagrangian level.
By working in the Hamiltonian formalism and starting from an equivalent description of the model as a two-parameter deformation of the coset $\sigma$-model on $G \times G / \Gd$, we show that it also admits a Lax matrix whose Poisson bracket is of the standard $r/s$-form characterised by a twist function which we determine. A number of results immediately follow from this, including the identification of certain complex Poisson commuting Kac-Moody currents as well as an explicit description of the $q$-deformed symmetries of the model. Moreover, the model is also shown to fit naturally in the general scheme recently developed for constructing integrable deformations of $\sigma$-models. Finally, we show that although the Poisson bracket of the Lax matrix still takes the $r/s$-form after fixing the $\Gd$ gauge symmetry, it is no longer characterised by a twist function.
\end{abstract}

\setcounter{tocdepth}{2}
\tableofcontents

\section{Introduction}

The Yang-Baxter $\sigma$-model is a one-parameter deformation of the principal chiral model, first introduced by C. {Klim$\check{\text{c}}$\'{\i}k} more than twenty years ago \cite{Klimcik:2002zj}. Its name stems from the presence of a solution of the modified classical Yang-Baxter equation in its action. 
The classical integrability of this model at the Lagrangian level was later proved in \cite{Klimcik:2008eq} 
by exhibiting a Lax pair, the flatness of which reproduces the equations of motion. Recently, the 
Yang-Baxter $\sigma$-model was recovered as the simplest case of a general procedure developed 
to deform a broad class of integrable $\sigma$-models while preserving their integrability 
\cite{Delduc:2013fga,Delduc:2013qra}. 
The whole construction is deeply rooted in the Hamiltonian formalism. In particular, one of its salient features is that the integrability at the Hamiltonian level of the resulting deformed $\sigma$-models is ensured from the very outset.

\medskip

Recall that proving Hamiltonian integrability requires more than determining a Lax pair. Indeed, the existence of a Lax pair only implies that there is an infinite number of conserved quantities. However, the Hamiltonian definition of integrability requires showing instead that there is an infinite number of 
quantities Poisson commuting with one another, not just with the Hamiltonian. Such a property is guaranteed if the Poisson bracket of the Lax matrix, defined as the spatial component of the Lax pair, 
can be put in the general $r/s$-form \cite{Maillet:1985fn, Maillet:1985ek}. Furthermore, it was shown in \cite{Sevostyanov:1995hd} for the principal chiral model, and in \cite{Vicedo:2010qd} for symmetric space $\sigma$-models and the $AdS_5 \times S^5$ superstring theory, that the algebraic structure behind the $r/s$-form of these $\sigma$-models is encoded in a so called twist function.

\medskip

The twist function of a given integrable $\sigma$-model plays a key role in the study of its integrable deformations. Indeed, the one-parameter integrable deformations of the principal chiral model and (semi-)symmetric $\sigma$-models constructed in \cite{Delduc:2013fga,Delduc:2014kha} were obtained by deforming their twist functions. More precisely, the focus of \cite{Delduc:2013fga,Delduc:2014kha} was on the so called Yang-Baxter class of deformations, of which the Yang-Baxter $\sigma$-model is the prototype. There exists another way of deforming the 
$\sigma$-models in question, with a completely different Lagrangian description 
\cite{Sfetsos:2013wia,Hollowood:2014rla,Hollowood:2014qma,Itsios:2014vfa,
Sfetsos:2014cea,
Sfetsos:2014lla,
Demulder:2015lva,
Sfetsos:2015nya,
Hollowood:2015dpa,
Appadu:2015nfa}. Nevertheless, in the
Hamiltonian framework, the procedure for obtaining these alternative deformations may also 
be interpreted as deforming the corresponding twist functions 
\cite{Hollowood:2014rla,Vicedo:2015pna}.
For completeness, let us also mention that within the Yang-Baxter class of integrable 
deformations there is also a way to deform a given $\sigma$-model by using a 
solution of the classical Yang-Baxter equation 
\cite{Kawaguchi:2014qwa, Kawaguchi:2014fca, 
Matsumoto:2014nra, Matsumoto:2014gwa,
Crichigno:2014ipa,Matsumoto:2014ubv,
   Matsumoto:2015jja,Matsumoto:2015uja,Matsumoto:2014cja,
   vanTongeren:2015soa,vanTongeren:2015uha}, but without changing its 
twist function \cite{Vicedo:2015pna}.

\medskip

The bi-Yang-Baxter $\sigma$-model was also proposed in \cite{Klimcik:2008eq} as a two-parameter deformation of the principal chiral model. Its Lagrangian integrability was only proved relatively recently in \cite{Klimcik:2014bta}. 
An interesting feature of this model is the following. Whereas the principal chiral model on a real Lie group $G$ admits an invariance under $G \times G$ by left and right multiplications of the $G$-valued field, in the Yang-Baxter $\sigma$-model one of these two global symmetries gets 
 deformed to $U^{\mathcal P}_q(\g)$, the Poisson algebra analogue of a quantum group. Here $q$ is a function of the single deformation parameter. The bi-Yang-Baxter $\sigma$-model can be seen as a further deformation of the Yang-Baxter $\sigma$-model in which both left and right 
global $G$-symmetries get deformed \cite{Klimcik:2008eq}.

\medskip

In this article we will focus on the Hamiltonian analysis of the bi-Yang-Baxter $\sigma$-model. In section \ref{sec-2}, we begin by recalling the action of the bi-Yang-Baxter $\sigma$-model. We start from its formulation as a two-parameter deformation of the coset $\sigma$-model on $G \times G/\Gd$, where $\Gd$ is the diagonal subgroup of $G \times G$. That is, when both deformation parameters are turned off we obtain the coset $\sigma$-model on $G \times G/\Gd$. The principal chiral model on $G$ is then recovered in a particular gauge. This point of view on the bi-Yang-Baxter $\sigma$-model was recently adopted in \cite{Hoare:2014oua} where the corresponding Lax pair was introduced. 
Since the deformation preserves the gauge invariance under $\Gd$, a first-class constraint appears in the canonical analysis. 
In the presence of such constraints, the Hamiltonian Lax matrix $\L(z)$, with $z$ the spectral parameter, is not fully determined by its Lagrangian counterpart. Indeed, one has the freedom to add to the latter a term consisting of an arbitrary function $f(z)$ times the constraint. This freedom was first shown to play an important role in \cite{Magro:2008dv,Vicedo:2009sn} for the $AdS_5 \times S^5$ superstring theory. 

In section \ref{sec-3} we show that the Poisson bracket of $\L(z)$ and $\L(z')$ takes 
the desired $r/s$-form ensuring Hamiltonian integrability for a specific  
choice of the function $f(z)$. More precisely, since we are considering a deformation of the coset $\sigma$-model on $G \times G/\Gd$, the Lax matrix naturally takes values in the twisted loop algebra of the real double $D\g = \g \oplus \g$ of the Lie algebra $\g$ of $G$. However, in this particular case it is possible to work instead with a Lax matrix taking values in the loop algebra of a single copy of $\g$. The corresponding $r$- and $s$-matrices are the skew-symmetric and symmetric parts, respectively, of an ${\cal R}$-matrix of the standard form depending on a two-parameter twist function $\varphi_\bYB(z)$ which we determine.

To complete the analysis, in section \ref{DoubleAlgebra} we indicate how the result obtained may be understood when working with a Lax matrix valued in the twisted loop algebra of $D\g$. In this formalism, the Poisson bracket of the Lax matrix with itself is still of the $r/s$-form but where the ${\cal R}$-matrix takes on a novel form depending on both the twist function $\varphi_\bYB(z)$ and its ``mirror'' image $\varphi_\bYB(-z)$. This ${\cal R}$-matrix is shown to correspond to the kernel of the standard solution of the modified classical Yang-Baxter equation on the twisted loop algebra of $D\g$ but with respect to an non-standard inner product on the latter. All these results show that  the bi-Yang-Baxter $\sigma$-model belongs to the same class of deformations as those constructed in \cite{Delduc:2013fga}. Indeed, it corresponds to a deformation of the twist function of the $G \times G/\Gd$ coset $\sigma$-model. 

In section \ref{sec-lift}, we recall the importance of studying the poles of the twist function. 
Specifically, we show that the Lax matrix $\L(z)$ evaluated at the poles of the twist function 
$\varphi_\bYB(z)$ yields a pair of Poisson commuting Kac-Moody currents valued in the 
complexification $\g^\CC = \g \otimes \CC$ of the real Lie algebra $\g$. We go on to show 
how the canonical fields of the bi-Yang-Baxter $\sigma$-model 
  may be recovered from the Lax matrix at the poles of the twist function. The upshot of this analysis 
  is that the bi-Yang-Baxter $\sigma$-model also fits the general scheme described in 
  \cite{Vicedo:2015pna}. As another important output of studying the (gauge transformed) 
  monodromy matrix at the poles of $\varphi_\bYB(z)$, it immediately follows that the 
  global $G \times G$ symmetry of the principal chiral model gets deformed to 
  $U^\P_q(\g) \times U^\P_{\tilde{q}}(\g)$. We indicate how we recover the values of 
  $q$ and $\tilde{q}$ first given in \cite{Hoare:2014oua}. This generalises the situation in \cite{Delduc:2013fga} recalled above, and which first appeared in the context of the Yang-Baxter $\sigma$-model on $SU(2)$, also known as the squashed $S^3$ $\sigma$-model \cite{Kawaguchi:2011pf, Kawaguchi:2012gp}.
 
Finally, in section \ref{sec-gf} we study the fate of the $r/s$-form of the Lax 
matrix algebra when gauge fixing the local $\Gd$-symmetry of the bi-Yang-Baxter 
$\sigma$-model. We do this by regarding the gauge fixing as a gauge 
transformation on 
the Lax matrix. 
This enables one to determine 
how the $r/s$-form behaves under  this gauge fixing. We show that the $r$- and 
$s$-matrices are no longer fully determined by a twist function but depend 
also on the 
$R$-matrices characterising the Yang-Baxter type deformation.

\section{The bi-Yang-Baxter $\sigma$-model} 
\label{sec-2}
\subsection{Lagrangian analysis}

\subsubsection{Action}

Let $G$ be a semi-simple real Lie group with Lie algebra  $\g$. 
Let $R$ and $\Rt$ be two skew-symmetric solutions of the modified classical Yang-Baxter 
equation (mCYBE) on $\g$, \textit{i.e.} endomorphisms of $\g$ such that for every $x,y \in \g$, we have
\begin{subequations}
\begin{align}
\kappa(x, Ry) &= - \kappa(Rx, y), \label{RSkewSym} \\
\left[Rx,Ry\right] &= R\bigl([Rx,y]+[x,Ry]\bigr)+[x,y], \label{mCYBE} 
\end{align}
\end{subequations}
and similarly for $\Rt$. Here $\kappa$ denotes the Killing form on $\g$ defined as $\kappa(x, y) = -\tr \big( \text{ad}_x \text{ad}_y \big)$ for any $x, y \in \g$.

We then consider the bi-Yang-Baxter $\sigma$-model associated with $R$ and $\Rt$, defined by the following action for a field $(g,\gt)$ valued in the double group $G\times G$ 
\cite{Hoare:2014oua}
\begin{equation}\label{Action}
S[g,\gt]= K \int d\tau d\s \; \kappa\left(j_+-\jt_+,\left(1-\frac{\eta}{2} R_g-\frac{\etat}{2}\Rt_{\gt} 
\right)^{-1} (j_--\jt_-)\right).
\end{equation}
 $K$, $\eta$ and $\etat$ are real parameters, $\p_\pm = \p_\tau \pm \p_\s$, 
 and   we have introduced the following notations 
\begin{gather*}
j_\pm = g^{-1}\p_\pm g,  \qquad 
\jt_\pm = \gt^{-1}\p_\pm \gt, \\
R_g = \Ad_g^{-1} \circ R \circ \Ad_g, \qquad 
\Rt_{\gt} = \Ad_{\gt}^{-1} \circ \Rt \circ \Ad_{\gt}, \\
\Ad_g(M) = gMg^{-1}.
\end{gather*}
Let us notice here that $R_g$ and $\Rt_{\gt}$ are also skew-symmetric solutions of the mCYBE.\\

When $\eta=\etat=0$ we recover the coset $\sigma$-model on the quotient $G\times G/\Gd$ by the diagonal subgroup $\Gd$ of $G \times G$.  It is direct to check that, like the coset $\sigma$-model, the bi-Yang-Baxter $\sigma$-model is invariant 
under gauge transformations taking values in the subgroup $\Gd$, namely
\begin{equation}\label{GaugeSym}
g \mapsto gh^{-1} \; \; \; \; \; \text{and} \; \; \; \; \; \gt \mapsto \gt h^{-1},
\end{equation}
with $h$ a field valued in the group $G$. We may impose the gauge fixing condition $\gt=\Id$, which 
is attained by performing the gauge transformation \eqref{GaugeSym} with $h=\gt$. This leads 
to a model for the $G$-valued field $g'=g\gt^{-1}$, which coincides with the two-parameter deformation of the principal chiral model first introduced in \cite{Klimcik:2008eq}.

\subsubsection{Equations of motion}

The equation of motion for the field $g$ derived from the action \eqref{Action} can be written as
\begin{equation}\label{EOM}
EOM=\p_+J_-+[a_+,J_-] + \p_-J_++[a_-,J_+] = 0,
\end{equation}
where we introduced
\begin{equation}\label{DefJ}
J_\pm = \left( 1\pm\dfrac{\eta}{2}R_g\pm\dfrac{\etat}{2}\Rt_{\gt} \right)^{-1} (j_\pm-\jt_\pm)
\end{equation}
and a ``gauge field''
\begin{equation}\label{Defa}
a_\pm = j_\pm \mp \frac{\eta}{2} R_gJ_\pm = \left(1\pm \frac{\etat}{2}\Rt_{\gt}\right)J_\pm+\jt_\pm .
\end{equation}

\noi Notice that a transformation
\begin{equation}\label{aFreedom}
a_\pm \mapsto a_\pm + \alpha J_\pm
\end{equation}
of the gauge field does not change the equation of motion \eqref{EOM}. 

The action \eqref{Action} is not changed when one exchanges $\eta$, $R$ and $g$ with $\etat$, $\Rt$ and $\gt$. Thus the equation of motion for $\gt$ takes the same form:
\beqz
\widetilde{EOM}=\p_+\Jt_-+[\at_+,\Jt_-] + \p_-\Jt_++[\at_-,\Jt_+] = 0,
\eeqz
with
\begin{align*}
\Jt_\pm &= \left( 1 \pm\dfrac{\eta}{2}R_g\pm\dfrac{\etat}{2}\Rt_{\gt} \right)^{-1}(\jt_\pm-j_\pm),\\
\at_\pm &= \jt_\pm \mp \frac{\etat}{2} \Rt_{\gt}\Jt_\pm = \left( 1\pm \frac{\eta}{2} R_g \right)\Jt_\pm+j_\pm.
\end{align*}

\noi It is then easy to check that
\begin{equation}\label{JaTilde}
\Jt_\pm = - J_\pm \; \; \; \; \; \; \text{and} \; \; \; \; \; \; \at_\pm = a_\pm - J_\pm .
\end{equation}
Thus, using the freedom \eqref{aFreedom} on $a_\pm$, we see that $\widetilde{EOM} = -EOM$. Therefore, the equation of motion for $\gt$ is equivalent to the one for $g$.

\subsubsection{Lax pair}

In this subsection, we recall that the equation of motion \eqref{EOM} can be cast in the form of a zero curvature equation
\begin{equation}\label{LaxEq}
\p_+\Lc_-(z)-\p_-\Lc_+(z)-[\Lc_+(z),\Lc_-(z)]=0
\end{equation}
for a Lax pair $\Lc_\pm(z)$ depending on a spectral parameter 
$z$~\cite{Hoare:2014oua}. Starting from the Maurer-Cartan equation on $j_\pm$, 
\beqz
 \p_+j_- - \p_-j_+ + [j_+,j_-] = 0,
\eeqz
we re-express it in terms of $J_\pm$ and $a_\pm$ using \eqref{Defa}, giving
\begin{equation}\label{MC}
 \p_+a_- - \p_-a_+ + [a_+,a_-] + \frac{\eta^2}{4} [J_+,J_-] - \frac{\eta}{2} R_g(EOM) =0,
\end{equation}
where we used the mCYBE on $R_g$. In the same way, the Maurer-Cartan equation on 
$\jt_\pm$ reads 
\beq \label{dec2a}
  \p_+\at_- - \p_-\at_+ + [\at_+,\at_-] + \frac{\etat^2}{4} [J_+,J_-] + 
\frac{\etat}{2} \Rt_{\gt}(EOM) =0,
\eeq
where we have used $\Jt_\pm=-J_\pm$ and $\widetilde{EOM}=-EOM$. 
Taking the difference between \eqref{MC} and  \eqref{dec2a} 
and using \eqref{JaTilde}, we obtain
\begin{equation}\label{DiffMC}
\p_+J_- + [a_+,J_-] - \p_-J_+ - [a_-,J_+] - \left(1-\frac{\eta^2-\etat^2}{4}\right) [J_+,J_-] 
- \frac{1}{2}(\eta R_g + \etat\Rt_{\gt})(EOM) =0.
\end{equation}
We introduce new gauge fields
\begin{equation}\label{DefA}
A_\pm = a_\pm - \frac{1}{2}\left(1-\frac{\eta^2-\etat^2}{4} \right)J_\pm.
\end{equation}
In terms of these, the equation of motion \eqref{EOM} keeps the same form
\begin{equation}\label{EOM2}
EOM=\p_+J_-+[A_+,J_-] + \p_-J_++[A_-,J_+]
\end{equation}
and the equation \eqref{DiffMC} becomes
\begin{equation}\label{DiffMC2}
 \p_+J_- + [A_+,J_-] - \p_-J_+ - [A_-,J_+] - \ha (\eta R_g + \etat\Rt_{\gt})(EOM)=0.
\end{equation}\vspace{4pt}
Coming back to the expression \eqref{MC} and using the definition \eqref{DefA} of 
$A_\pm$, we find  
\begin{multline}
0 = \p_+A_- - \p_-A_+ + [A_+,A_-] +\frac{\zeta^2}{4} [J_+,J_-] \\ 
+ \qa\Bigl(1-\frac{\eta^2-\etat^2}{4} \Bigr) \bigl(\eta R_g + \etat\Rt_{\gt}\bigr)(EOM) - \eta R_g(EOM) \label{dec2b}
\end{multline}
where 
\beq \label{def-zeta}
\zeta = \sqrt{\bigl(1+\qa (\eta+\etat)^2\bigr)\bigl(1+\qa(\eta-\etat)^2\bigr)}.
\eeq
Finally, taking the equation \eqref{dec2b} on shell ($EOM=0$) and the sum and the difference 
of equations \eqref{EOM2} and \eqref{DiffMC2} also on shell, we arrive at 
\begin{gather*}
 \p_+A_- - \p_-A_+ + [A_+,A_-] + \frac{\zeta^2}{4} [J_+,J_-]=0, \\
  \p_+J_- + [A_+,J_-]=0  \et 
  \p_-J_+ + [A_-,J_+]=0.
  \end{gather*}

\noi It is easy to see that these three equations are equivalent to the zero curvature equation \eqref{LaxEq} for the Lax pair:
\begin{equation}\label{LaxPair}
\Lc_\pm(z) = -A_\pm - \frac{\zeta}{2} z^{\pm 1} J_\pm .
\end{equation}

\subsection{Hamiltonian analysis}

\subsubsection{Conjugate momentum}

Let us introduce a basis $T_a$ of the Lie algebra $\g$ and coordinates $\phi^i$ on the group $G$. We denote $\p_i$ the derivation with respect to the coordinate $\phi^i$. We can then introduce $L_i^a$ such that
\beqz
g^{-1} \p_i g = L_i^a T_a.
\eeqz
\noi From the action \eqref{Action}, we compute the conjugate momenta $\pi_i$ of the coordinates $\phi^i$ to be
\begin{align*}
\pi_i = K L^a_i &\left[\kappa\left(T_a,\left(1-\frac{\eta}{2} R_g-\frac{\etat}{2}\Rt_{\gt} 
\right)^{-1}(j_--\jt_-)\right) \right. \notag\\
&\qquad\qquad\qquad +\left. \kappa\left(j_+-\jt_+,\left(1-\frac{\eta}{2} R_g-
\frac{\etat}{2}\Rt_{\gt} \right)^{-1}T_a\right) \right].
\end{align*}

\noi Using the skew-symmetry of $R$ and \eqref{DefJ}, we have 
\vspace{-2pt}
\begin{equation}\label{ExpressionPi}
\pi_i = K L^a_i  \kappa(T_a, J_-  +   J_+).
\end{equation}
with the metric $\kappa_{ab}=\kappa(T_a,T_b)$. It is more convenient to introduce the following $\g$-valued field
\begin{equation}\label{DefX}
X=L^i_a\pi_i\kappa^{ab}T_b,
\end{equation}
where $L^i_a$ is the inverse of $L_i^a$ and $\kappa^{ab}$ is the inverse of the metric $\kappa_{ab}$. In particular, one can check that these fields are independent of the choice of coordinates $\phi^i$ and of basis $T_a$. It is then easy to deduce the expression of $X$ from \eqref{ExpressionPi} to be
\begin{equation} \label{X vs Jpm}
X=K(J_++J_-).
\end{equation}
In the same way, one would find $\Xt=K(\Jt_++\Jt_-)=-K(J_++J_-)$. Thus, we have the constraint
\begin{equation}\label{Constraint}
X+\Xt=0.
\end{equation}
This is a consequence of the gauge symmetry \eqref{GaugeSym} of the model.

\subsubsection{Poisson brackets and Hamiltonian density}

We start with the canonical Poisson brackets
\begin{equation}
\lbrace \pi_i(\s), \phi^j(\s') \rbrace = \delta_i^j \delta_{\s\s'}.
\end{equation}
where $\delta_{\s\s'}$ is the Dirac $\delta$-distribution. From those canonical Poisson brackets and the definition \eqref{DefX} of $X$, we deduce 
the classical brackets on the fields $g$ and $X$ parametrising the cotangent 
bundle $T^\ast LG$, with $LG$ the loop group associated with $G$, to be
\begin{subequations}\label{PBgX}
\begin{align}
\left\lbrace g\ti{1}(\s), g\ti{2}(\s') \right\rbrace & =  0, \\
\left\lbrace g\ti{1}(\s), X\ti{2}(\s') \right\rbrace & =  - g\ti{1}(\s) C\ti{12} \delta_{\s\s'}, \\
\left\lbrace X\ti{1}(\s), X\ti{2}(\s') \right\rbrace & =  - \left[ C\ti{12}, X\ti{2}(\s) \right] \delta_{\s\s'}. \label{PBXX}
\end{align}
\end{subequations}
We used  standard tensorial notations with subscripts $\underline{1}$ and $\underline{2}$ and $C_{\1\2}=\kappa^{ab} T_a \otimes T_b$ is the split Casimir. The fields $\gt$ and $\Xt$ parametrising another copy of $T^\ast LG$ verify the same Poisson brackets. All other brackets vanish. Moreover, as long as we are calculating Poisson brackets, we must consider $X$ and $\Xt$ as independent variables in the phase space, without imposing the constraint \eqref{Constraint}.

\medskip

The Legendre transform of the Lagrangian in \eqref{Action} is the ``naive'' Hamiltonian density
\begin{equation}\label{NaiveHam}
\mathcal{H}_0 = \frac{K}{2} \bigl( \kappa \left( J_+,J_+ \right) + \kappa \left( J_-,J_- \right) \bigr).
\end{equation}

As we are considering a constrained system, we have to follow the Dirac procedure and add a term proportional to the constraint to define the Hamiltonian density of the system
\begin{equation}\label{Ham}
\mathcal{H} = \mathcal{H}_0 + \kappa \bigl( \Lambda, X+\Xt \bigr),
\end{equation}
where $\Lambda$ is a $\g$-valued field playing the role of a Lagrange multiplier. 
There is no secondary constraint. 

\subsubsection{Hamiltonian Lax matrix}

Let us now determine the form of the Hamiltonian Lax matrix of the model.
At the Lagrangian level, the Lax matrix is given by the spatial component of the Lax pair, \textit{i.e.} by $\frac{1}{2}(\Lc_+-\Lc_-)$. As we are considering a constrained Hamiltonian system, we have the freedom of adding a term proportional to the constraint, thus getting
\beqz
\Lc(z) = \frac{1}{2}\big( \Lc_+(z)-\Lc_-(z) \big) + f(z)(X+\Xt),
\eeqz
where $f$ is some function of $z$, which will be fixed later to ensure the Hamiltonian integrability of the model. One 
could potentially add other extra terms, for instance 
proportional to $R_g(X+\Xt)$ and $\Rt_{\gt}(X+\Xt)$, but as we will see in the next section, they turn out not to be necessary.

Using equations \eqref{LaxPair} and \eqref{DefA}, we get
\beqz
\Lc(z) = -\frac{1}{2}(a_+-a_-) + \frac{1}{4}\left(1-\frac{\eta^2-\etat^2}{4} \right) (J_+-J_-) - \frac{\zeta}{4} \left( zJ_+ - \frac{1}{z}J_- \right) + f(z)(X+\Xt).
\eeqz

\noi The definition \eqref{Defa} of $a_\pm$ can be re-written in a more symmetric way as
\beqz
a_\pm  = \frac{1}{2}\left(j_\pm + \jt_\pm + J_\pm \mp \frac{\eta}{2} R_gJ_\pm
\pm \frac{\etat}{2}\Rt_{\gt}J_\pm \right),
\eeqz
thus giving
\beqz
a_+-a_- = \frac{1}{2}\left(j_+-j_- + \jt_+-\jt_- + J_+-J_- - \left(\frac{\eta}{2} R_g -
\frac{\etat}{2} \Rt_{\gt} \right) (J_++J_-) \right).
\eeqz
Denoting $j=\frac{1}{2}(j_+-j_-)$ and $\jt=\frac{1}{2}(\jt_+-\jt_-)$, we obtain
\begin{multline*}
\Lc(z) = -\frac{1}{2}(j+\jt) - \left(\frac{\eta^2-\etat^2}{16}+
\frac{\zeta}{8}\left(z+\frac{1}{z}\right)\right)(J_+-J_-) - \frac{\zeta}{8} \left(z-\frac{1}{z}\right) (J_++J_-) \\ 
+ \left(\frac{\eta}{8} R_g - \frac{\etat}{8} \Rt_{\gt} \right) (J_++J_-) + f(z)(X+\Xt). 
\end{multline*} 
Using \eqref{DefJ}, we have
\beqz
J_+-J_- = 2j - 2\jt - \left(\frac{\eta}{2}R_g + \frac{\etat}{2}\Rt_{\gt} \right) (J_++J_-),
\eeqz
which gives
\begin{multline} \label{Lax Ham pre}
\Lc(z) = -\frac{1}{2}\left(1+\frac{\eta^2-\etat^2}{4}+\frac{\zeta}{2}\left(z+\frac{1}{z}\right)\right)j-\frac{1}{2}\left(1+\frac{\etat^2-\eta^2}{4}-\frac{\zeta}{2}\left(z+\frac{1}{z}\right)\right)\jt \\
+\frac{\eta}{8}\left(1 + \frac{\eta^2-\etat^2}{4}+\frac{\zeta}{2}\left(z+\frac{1}{z}\right)\right)R_g(J_++J_-)\\
\qquad\qquad\qquad - \frac{\etat}{8}\left(1 + \frac{\etat^2-\eta^2}{4}-\frac{\zeta}{2}\left(z+\frac{1}{z}\right)\right)\Rt_{\gt}(J_++J_-) \\
-\frac{\zeta}{8}\left(z-\frac{1}{z}\right)(J_++J_-) + f(z) (X+\Xt).
\end{multline}
In order to finish re-expressing \eqref{Lax Ham pre} in terms of the Hamiltonian 
fields alone, we make use of equations \eqref{X vs Jpm} and \eqref{Constraint} namely $J_++J_-
=X/K=-\Xt/K$. For reasons of symmetry and simplicity, we will 
use $X$ (respectively $\Xt$) when $R_g$ (respectively $\Rt_{\gt}$) is applied to 
$J_++J_-$, and we will use the linear combination $\ha(X-\Xt)$ when $J_++J_-$ stands alone. 
This last ``prescription'' does not change the expression of the Lax matrix, as any other 
choice can be re-absorbed in the function $f(z)$ which is so far arbitrary. Beyond the 
arguments of symmetry, the resulting form of the Hamiltonian Lax matrix will be 
justified in the following section to prove the Hamiltonian integrability of the model.

The final result can be written in terms of the set of fields $\Ob=\lbrace j,X,R_gX,\jt,\Xt,\RgXt\rbrace$ as
\begin{equation}\label{Lax}
\Lc(z) = \sum_{Q\in\Ob} A_Q(z) Q,
\end{equation}
with coefficients $A_Q$ whose expressions are given in appendix \ref{CoeffJ}.

\subsection{One-parameter deformation limit}
\label{OneParam}

By fixing $\eta=\etat$ we obtain a one-parameter deformation of the coset model on $G\times G / \Gd$. It is given by the action
\begin{equation} \label{dec2c}
S[g,\gt]= K \int d\s d\tau \; \kappa\left(j_+-\jt_+,\left(1-\frac{\eta}{2} R_g-\frac{\eta}{2}\Rt_{\gt} \right)^{-1} (j_--\jt_-)\right).
\end{equation}

Let us consider the double Lie group $DG=G\times G$ and the corresponding double Lie algebra $D\g=\g\oplus\g$. The latter comes naturally equipped with the exchange automorphism
\begin{equation}\label{DefDelta}
\begin{array}{rccc}
\delta: & D\g & \longrightarrow & D\g \\
        & (x,y) & \longmapsto   & (y,x)
\end{array}.
\end{equation}
We may decompose $D\g$ into eigenspaces of this involution as 
 $D\g = D\g^{(0)} \oplus D\g^{(1)}$, 
with $D\g^{(0)} = \ker(\delta-\Id)$ and $D\g^{(1)} = \ker(\delta+\Id)$. 
We can notice here that $D\g^{(0)}=\g_{\text{diag}}$, the Lie algebra of the diagonal subgroup $G_{\text{diag}}$, so that the quotient $G\times G / \Gd$ is indeed the coset $DG/DG^{(0)}$.

We will denote $P_0$ and $P_1$ the projectors associated with this decomposition, defined by
\begin{equation}
\begin{array}{rccccccrccc}
P_0 : & D\g & \longrightarrow & D\g & \; \; \; \; & \text{and} & \; \; \; \; & P_1 : & D\g & \longrightarrow & D\g \\
      & (x,y) & \longmapsto & \frac{1}{2}(x+y,x+y) & & & & & (x,y) & \longmapsto & \frac{1}{2}(x-y,y-x)
\end{array}
\end{equation}
In this formulation on the double Lie group and Lie algebra, it is natural to introduce the 
field $h=(g,\gt) \in DG$ and the solution $\mathfrak{R}=(R,\Rt) \in \text{End}(D\g)$ of 
the mCYBE on $D\g$. The action \eqref{dec2c} can then be re-expressed as
\begin{equation}\label{ActionOneParam}
S[g,\gt]= 2K \int d\s d\tau \; \kappa\left((h^{-1}\p_+h)^{(1)},\left(1-\eta \, \mathfrak{R}_h \circ P_1 \right)^{-1} (h^{-1}\p_-h)^{(1)}\right).
\end{equation}
This is nothing but the one-parameter deformation of the coset $\sigma$-model 
introduced in \cite{Delduc:2013fga} when the quotient considered is $G\times G/\Gd$ and 
with $K=\frac{1}{4}(1+\eta^2)$.

\section{Hamiltonian integrability}
\label{sec-3}
In this section, we will compute the Poisson bracket of the Lax matrix \eqref{Lax} with 
itself and show that it can be cast in the $r/s$-form (more precisely an $r/s$-system 
involving twist function), thus proving the Hamiltonian integrability of the 
bi-Yang-Baxter $\sigma$-model. 

\subsection{$r/s$-form and twist functions}

Let $\Rc_{\1\2}(z,z')$ be a rational function of $z$ and $z'$ valued in $\g^\CC \otimes \g^\CC$, where $\g^\CC$ is the complexification of $\g$, and satisfying the classical Yang-Baxter equation with spectral parameters. We do not assume that $\Rc_{\1\2}(z, z')$ is skew-symmetric, \emph{i.e.} that it has the property $\Rc_{\1\2}(z, z') = - \Rc_{\2\1}(z', z)$. We introduce its skew-symmetric and symmetric parts as
\begin{subequations}\label{SystemRS}
\begin{equation}\label{RandS}
r_{\1\2}(z,z') = \frac{1}{2}\bigl(\Rc_{\1\2}(z,z') - \Rc_{\2\1}(z',z) \bigr) \; \; \; \; \text{and} \; \; \; \; 
s_{\1\2}(z,z') = \frac{1}{2}\bigl(\Rc_{\1\2}(z,z') + \Rc_{\2\1}(z',z) \bigr).
\end{equation}
The Poisson bracket of the Lax matrix with itself is said to be of the $r/s$-form, 
associated with this matrix $\Rc$, if it can be written as \cite{Maillet:1985fn,Maillet:1985ek}
\begin{align}\label{RS}
\left\lbrace \Lc\ti{1}(z,\sigma), \Lc\ti{2}(z',\sigma') \right\rbrace &=
\left[ r\ti{12}(z,z'), \Lc\ti{1}(z,\s)+\Lc\ti{2}(z',\s') \right] \delta_{\s\s'} \notag\\
 &\qquad + \left[ s\ti{12}(z,z'), \Lc\ti{1}(z,\s)-\Lc\ti{2}(z',\s') \right] \delta_{\s\s'} + \; 2 s\ti{12}(z,z') \delta'_{\s\s'},
\end{align}
where $\delta'_{\s\s'} = \partial_\s \delta_{\s\s'}$.  

The non-ultralocality of this Poisson bracket, namely the presence of $\delta'$-terms, is completely characterised by the symmetric part of the $\Rc$-matrix being non-zero. For a very broad class of integrable $\sigma$-models, the $\Rc$-matrix $\Rc_{\1\2}(z, z')$ is given by the kernel of an abstract solution of the mCYBE on the loop algebra $\g(\!( z )\!)$, with respect to the standard inner product on $\g(\!( z )\!)$ modified by a rational function $\varphi(z)$, called the twist function (see for instance \cite{Vicedo:2010qd} or section \ref{sec: R-matrix} for the case when $\g$ is replaced by the double $D\g$). In this situation the failure of $\Rc$ to be skew-symmetric is encoded in the twist function. In the simplest of cases, the kernel $\Rc_{\1\2}(z,z')$ takes the form
\vspace{-5pt}\begin{equation}\label{RSTwist}
\Rc_{\1\2}(z,z')=\frac{C\ti{12}}{z-z'}\varphi(z')^{-1},
\end{equation}
\end{subequations}
and is therefore skew-symmetric if and only if $\varphi$ is constant.

The simplest example of a model with such an $\Rc$-matrix is the principal chiral model \cite{Sevostyanov:1995hd}. Moreover, one can show from the results of \cite{Delduc:2013fga} that the coset $\sigma$-model on $G\times G / \Gd$ and its one-parameter deformation also admit $\Rc$-matrices of this form\footnote{More precisely, \cite{Delduc:2013fga} deals with a general coset $\sigma$-model $F/G$. In the case of the coset $G\times G/\Gd$, we get a Lax matrix in the double algebra $D\g=\g\oplus\g$. We recover an $r/s$-system with an $\Rc$-matrix of the form \eqref{RSTwist} by taking the projector of this Lax matrix on the left part of $D\g$. This will be discussed in more details in section \ref{DoubleAlgebra} of the present article.}. The twist function of the coset $\sigma$-model on 
$G \times G/\Gd$ (which is, in the setting considered here, 
the limit $\eta=\etat=0$ of the bi-Yang-Baxter $\sigma$-model) is
\begin{equation}\label{TwistCoset}
\varphi_{\text{coset}}(z)=\frac{16Kz}{(1-z^2)^2}
\end{equation}
and the one of the Yang-Baxter deformation of this coset $\sigma$-model (which corresponds to $\eta=\etat$) is
\begin{equation}\label{TwistYB}
\varphi_{\text{YB}}(z)=\frac{16Kz}{(1 - z^2)^2 + \eta^2 (1 + z^2)^2}.
\end{equation}

We will now show that the bi-Yang-Baxter $\sigma$-model also admits an $\Rc$-matrix of the form \eqref{RSTwist} and will give the associated twist function.

\subsection{Expected form of the Poisson bracket}

We are seeking to put the Poisson bracket of the Lax matrix \eqref{Lax} in the $r/s$-form \eqref{SystemRS}, with a twist function $\varphi$ as in \eqref{RSTwist}. We will distinguish between two terms in this Poisson bracket: the ultralocal one, proportional to $\delta_{\s\s'}$, and the non-ultralocal one, proportional to $\delta'_{\s\s'}$. Let us write these as
\vspace{3pt}
\beqz
\left\lbrace \Lc\ti{1}(z,\sigma), \Lc\ti{2}(z',\sigma') \right\rbrace = P\ti{12}^{\text{UL}}(z,z',\s) 
\delta_{\s\s'} + P\ti{12}^{\text{NUL}}(z,z',\s) \delta'_{\s\s'}.
\eeqz

According to \eqref{RS}, the non-ultralocal term is directly proportional to the $s$-matrix. For a system with a twist function entering as in \eqref{RSTwist}, this term is thus given by
\begin{equation}\label{ExpectedNUL}
P_{\1\2}^{\text{NUL}}(z,z',\s) = -\frac{\varphi(z)^{-1}-\varphi(z')^{-1}}{z-z'}C_{\1\2}.
\end{equation}

The ultralocal term is slightly more complicated. Considering the expressions \eqref{RSTwist} of $\Rc$ and \eqref{Lax} of $\Lc$ and using the invariance property of the split Casimir, namely that for every $x\in\g$ we have $[C_{\1\2},x_\1+x_\2]=0$, one can reduce the ultralocal term to the form
\begin{equation}\label{ExpectedUL}
P\ti{12}^{\text{UL}}(z,z',\s) = \sum_{Q \in \Ob} J_Q(z,z') [C\ti{12},Q\ti{2}(\s)],
\end{equation}
with the coefficients $J_Q$ given by
\begin{equation}\label{ExpectedJ}
J_Q(z,z')=\frac{\varphi(z)^{-1} A_Q(z')-\varphi(z')^{-1} A_Q(z)}{z-z'}.
\end{equation}

\subsection{Poisson bracket of the Lax matrix}

We will now compute the Poisson bracket of the Lax matrix explicitly and compare the result to 
the expected form discussed in the previous subsection. Using equation \eqref{Lax}, this bracket is simply
\beqz
\left\lbrace \Lc\ti{1}(z,\s), \Lc\ti{2}(z',\s') \right\rbrace = 
\sum_{Q,Q'\in\Ob} A_Q(z) A_{Q'}(z') \lbrace Q\ti{1}(\s), Q'\ti{2}(\s') \rbrace .
\eeqz

The Poisson brackets between the different fields $Q\in\Ob=\lbrace j,X,R_gX,\jt,\Xt,\RgXt\rbrace$ 
can be derived from the basic Poisson brackets \eqref{PBgX}. In particular, let us mention 
that we have 
\beqz
\left\lbrace (R_gX)\ti{1}(\s), (R_gX)\ti{2}(\s') \right\rbrace = \left[ C\ti{12}, X\ti{2}(\s) \right] \delta_{\s\s'}.
\eeqz
This follows from the fact that $R_g$ is solution of the 
mCYBE. 
Any two fields from different copies of $\g$ Poisson commute.

\paragraph{Non-ultralocal term.}

The non-ultralocal term is generated by the brackets of $j$ and $\jt$ with the other fields. It reads
\begin{align*}
P\ti{12}^{\text{NUL}}(z,z',\s,\s') &= -\big(A_j(z)A_X(z')+A_j(z')A_X(z)+A_{\jt}(z)A_{\Xt}(z')+A_{\jt}(z')A_{\Xt}(z)\big)C\ti{12} \\
&\qquad + \big(A_j(z)A_{R_gX}(z')-A_j(z')A_{R_gX}(z)\big)R_g(\s)\ti{12}\\
&\qquad + \big(A_{\jt}(z)A_{\RgXt}(z')-A_{\jt}(z')A_{\RgXt}(z)\big)\Rt_{\gt}(\s)\ti{12},
\end{align*}
where we defined $R_g(\s)\ti{12}=R_{g(\s)\1}C\ti{12}$ and $\Rt_{\gt}(\s)\ti{12}=\Rt_{\gt(\s)\1}C\ti{12}$. One easily checks from \eqref{CoeffLax} that the coefficients of $R_g(\s)\ti{12}$ and $\Rt_{\gt}(\s)\ti{12}$ in this expression vanish. As expected in \eqref{ExpectedNUL}, we find a non-ultralocal term proportional to the split Casimir $C_{\1\2}$, namely
\begin{equation}\label{ActualNUL}
P\ti{12}^{\text{NUL}}(z,z',\s,\s') = -\left(A_j(z)A_X(z')+A_j(z')A_X(z)+A_{\jt}(z)A_{\Xt}(z')+A_{\jt}(z')A_{\Xt}(z)\right)C\ti{12}.
\end{equation}

\paragraph{Ultralocal term.}

We have in the ultralocal part three kinds of terms: 
\vspace{-3pt}\begin{itemize}
\setlength\itemsep{0.15em}
\item Terms proportional to $[C\ti{12},Q\ti{2}(\s)]$ with $Q\in\Ob$, as expected in \eqref{ExpectedUL}.
\item A term proportional to $[R_g(\s)\ti{12},j\ti{2}(\s)]$.
\item A term proportional to $[\Rt_{\gt}(\s)\ti{12},\jt\ti{2}(\s)]$.
\end{itemize}
The coefficients of the last two terms are the same as the coefficients of $R_g(\s)\ti{12}$ and $\Rt_{\gt}(\s)\ti{12}$ in the non-ultralocal term. Thus, they also vanish. We are then left with an ultralocal term of the form \eqref{ExpectedUL}. The expressions for the coefficients $J_Q(z,z')$ are given in appendix \ref{CoeffJ}.

\subsection{Twist function of the model}

To prove that the system admits a twist function, it remains to compare 
\eqref{ExpectedNUL} with \eqref{ActualNUL} and \eqref{ExpectedJ} with \eqref{ActualJd} 
and show that the different expressions match. We have shown that this is the case 
if we choose the function $f$ to be 
\beqz
f(z)= -\frac{\zeta^2}{16K}(1+z^2) + \frac{1}{8K}\left(1-\frac{(\eta^2-\etat^2)^2}{16}\right) 
- \frac{\zeta(\eta^2-\etat^2)}{64K}\left(3z+\frac{1}{z}\right),
\eeqz
where $\zeta$ is defined by equation \eqref{def-zeta}. 
The twist function is then
\begin{equation}\label{TwistFunction}
\varphi_\bYB(z)=\frac{1}{\zeta^2} \frac{16Kz}{z^4+\dfrac{\eta^2-\etat^2}
{\zeta}z^3+\left(2+\dfrac{(\eta^2-\etat^2)^2-16}{4\zeta^2}\right)z^2+\dfrac{\eta^2-\etat^2}{\zeta}z+1}.
\end{equation}
We will analyse the structure of this twist function in section \ref{sec-lift}.

\section{Formulation in the double Lie algebra}
\label{DoubleAlgebra}

Since we are considering a deformation of the coset $\sigma$-model on $G\times G / \Gd$, we would expect the Lax matrix to be valued in the twisted loop algebra of the real double $D\g=\g \oplus \g$, just as in the undeformed model \cite{Delduc:2012qb}. However, the Lax matrix discussed so far only takes values in the loop algebra of $\g$. We shall show in this section how the Hamiltonian integrability of the bi-Yang-Baxter $\sigma$-model can also be expressed using a formulation based on the double $D\g$.

\subsection{Lax pair in the double Lie algebra}

We will use the formalism of the double Lie algebra $D\g$ introduced in the 
subsection \ref{OneParam}. Let us consider the loop algebra associated with 
$D\g$, \textit{i.e.} the space $D\g (\!(z)\!) =D \g \otimes \CC(\!(z)\!)$ of Laurent series in a complex parameter $z$ valued in the complexification of $D\g$ and equipped with the natural Lie bracket. The exchange automorphism \eqref{DefDelta} on $D\g$ induces an automorphism $\hat{\delta}$ on $D\g(\!(z)\!)$ defined for all $X\in D\g(\!(z)\!)$ by
\begin{equation*}
\hat{\delta}(X)(z)=\delta\bigl(X(-z)\bigr).
\end{equation*}
Denote by $D\g(\!(z)\!)^{\hat{\delta}}$ the twisted loop algebra, \textit{i.e.} the 
subalgebra of $D\g(\!(z)\!)$ formed by the fixed points of $\hat{\delta}$.

Recall that the Lax matrices of the coset $\sigma$-model (corresponding 
 here to $\eta=\etat=0$) and of its one-parameter deformation (corresponding 
 here to $\eta=\etat$) belong to the twisted algebra $D\g(\!(z)\!)^{\hat{\delta}}$. It is 
 natural to expect such a Lax matrix to exist also for the bi-Yang-Baxter $\sigma$-model. 
  The corresponding Lax pair can be constructed 
  from the Lax pair $\Lc_\pm(z)$ valued in the loop algebra $\g(\!( z )\!)$ of a single copy of $\g$ in equation \eqref{LaxPair}. Indeed, defining
\beqz
L_\pm(z) = \bigl(\Lc_\pm(z),\Lc_\pm(-z)\bigr) \in D\g(\!( z )\!),
\eeqz
we have automatically $L _\pm(z) \in D\g(\!(z)\!)^{\hat{\delta}}$ and the Lax equation
\beqz
\p_+L_-(z)-\p_-L_-(z)-[L_+(z),L_-(z)]=0
\eeqz
follows immediately from the one for $\Lc_\pm(z)$ in \eqref{LaxEq}. The 
associated Hamiltonian Lax matrix is 
\beq \label{LaxDouble}
L(z) = \bigl(\Lc(z),\Lc(-z)\bigr) 
\eeq
where $\Lc(z)$ is given by \eqref{Lax}.

In the remainder of this section we will study the Hamiltonian properties of this Lax matrix, showing that its Poisson bracket is also of the $r/s$-form.

\subsection{Poisson bracket of the Lax matrix with itself}

The Lax matrices of the coset $\sigma$-model on $G\times G/\Gd$ and of its one-parameter deformation have a Poisson bracket of the $r/s$-form in the double algebra $D\g$. We will show that this is also the case for the bi-Yang-Baxter $\sigma$-model. As it turns out, however, the $\Rc$-matrix of the latter (which is a rational function of two spectral parameters $z$ and $z'$ valued in the complexification of $D\g \otimes D\g$) takes on a slightly non-standard form depending on both the twist function $\varphi_\bYB(z)$ and on its mirror image $\varphi_\bYB(-z)$. We will discuss the algebraic origin of this structure coming from the twisted loop algebra $D\g(\!( z )\!)^{\hat \delta}$ by generalising the construction of \cite{Vicedo:2010qd}. 

\subsubsection[$\Rc$-matrix and inner product]{$\bm{\Rc}$-matrix and inner product} \label{sec: R-matrix}

We begin by recalling the construction of \cite{Vicedo:2010qd} adapted to the present setting. The twisted loop algebra $D\g(\!(z)\!)$ admits a natural decomposition
\begin{equation} \label{Dg decomp}
D \g(\!(z)\!)^{\hat{\delta}} = D \g[[z]]^{\hat{\delta}} \oplus \left(z^{-1} D \g[z^{-1}]\right)^{\hat{\delta}}
\end{equation}
into subalgebras of positive and strictly negative powers of the loop parameter $z$, respectively. Let $\pi_+$ and $\pi_-$ denote the projection operators relative to this decomposition. The operator
\begin{equation} \label{ROp}
\Rc^D = \pi_+ - \pi_-
\end{equation}
defines a solution of the mCYBE on $D \g(\!(z)\!)^{\hat{\delta}}$.\\

Suppose now that we are given an invariant inner product $\langle \cdot, \cdot \rangle$ on the twisted loop algebra $D \g(\!(z)\!)^{\hat{\delta}}$. We define the kernel $\Rc^D\ti{12}(z, z')$ of the operator $\Rc^D$ in \eqref{ROp}, with respect to $\langle \cdot, \cdot \rangle$, as the rational function $\Rc^D_{\1\2}(z, z')$ of two complex variables and valued in the complexification of $D\g \otimes D\g$, such that for all $M \in D \g(\!(z)\!)^{\hat{\delta}}$ we have
\begin{equation}
\langle \Rc^D\ti{12}(z, z'),M_\2(z') \rangle \ti{2} = (\Rc^D M)(z).
\end{equation}
This matrix is then a solution of the classical Yang-Baxter equation\footnote{More precisely, it is a solution of the classical Yang-Baxter equation if we ignore contact terms 
by treating $\Rc^D_{\1\2}(z, z')$ as a rational function. See, for instance, \cite{Vicedo:2010qd} for more details.}
\begin{equation}
\left[\Rc^D\ti{12}(z_1,z_2),\Rc^D\ti{13}(z_1,z_3)\right]+\left[\Rc^D\ti{12}(z_1,z_2),\Rc^D\ti{23}(z_2,z_3)\right]+\left[\Rc^D\ti{32}(z_3,z_2),\Rc^D\ti{13}(z_1,z_3)\right] = 0.
\end{equation}
The standard inner product on $D\g(\!( z )\!)$ is defined for all $M, N \in D\g(\!(z)\!)$ by
\begin{equation}
\langle M, N \rangle = \text{res}_{z=0} \; \kappa^D \bigl( M(z),N(z) \bigr) dz,
\end{equation}
where $\kappa^D$ is the Killing form on the double $D\g$. Given any function $\varphi(z)$, one can also define a more general invariant inner product on $D\g(\!( z )\!)$ as a ``twist'' of the standard one by $\varphi$, namely
\begin{equation}\label{psOdd}
\langle M, N \rangle_\varphi = \text{res}_{z=0} \; \kappa^D \bigl( M(z),N(z) \bigr) \varphi(z) dz,
\end{equation}
for any $M, N \in D\g(\!(z)\!)$.
It is easy to check that this inner product is invariant under $\hat{\delta}$, \emph{i.e.} $\langle \hat \delta M, \hat \delta N \rangle_\varphi = \langle M, N \rangle_\varphi$, and thus induces an inner product on the twisted loop algebra $D\g(\!( z )\!)^{\hat \delta}$, if and only if $\varphi$ is an odd function. The kernel of the operator $\Rc^D$ defined in equation \eqref{ROp}, with respect to this inner-product, is
\begin{equation}\label{RTwistOdd}
\Rc^D\ti{12}(z, z') = 2 \frac{z' C^{(00)}\ti{12} + z C^{(11)}\ti{12}}{z'^2-z^2} \varphi(z')^{-1},
\end{equation}
with the graded components of the split Casimir 
\begin{subequations}
\begin{align}
C^{(00)}\ti{12} & =  \ha  \kappa^{ab}  \bigl( (T_a,0) + (0,T_a) \bigr) \otimes \bigl( (T_b,0) + (0,T_b) \bigr), \\
C^{(11)}\ti{12} & = \ha  \kappa^{ab}  \bigl( (T_a,0) - (0,T_a) \bigr) \otimes \bigl( (T_b,0) - (0,T_b) \bigr).
\end{align}
\end{subequations}
Expression \eqref{RTwistOdd} is the $\Rc$-matrix entering the $r/s$-form of the Poisson bracket of Lax matrices for the coset $\sigma$-model on $G\times G/\Gd$ as well as its one-parameter deformation, with the twist function $\varphi$ given respectively by \eqref{TwistCoset} and \eqref{TwistYB}.

\subsubsection{Inner product for the bi-Yang-Baxter $\sigma$-model}

Let us now generalise the ideas presented in the previous subsections, to have a formalism that also describes the bi-Yang-Baxter $\sigma$-model. As we are considering the double Lie algebra $D\g$, one can define an even more general inner product invariant under $\hat{\delta}$, by separating explicitly the left and right part of $D\g$. That is, for any $M=(m,\tilde{m})$ and $N=(n,\tilde{n})$ in $D\g$ we define
\begin{equation}\label{ps}
\langle M, N \rangle_\varphi = \text{res}_{z=0} \; \kappa \bigl( m(z),n(z) \bigr) \varphi(z) dz - \text{res}_{z=0} \; \kappa \bigl( \tilde{m}(z),\tilde{n}(z) \bigr) \varphi(-z) dz,
\end{equation}
where $\kappa$ is the Killing form on $\g$.
When $\varphi$ is odd, we recover the twisted inner product \eqref{psOdd}. This construction allows to consider twist functions of any parity.

The kernel of $\Rc^D$ with respect to the inner product \eqref{ps} is given by
\begin{equation}\label{RDouble}
\Rc^D_{\1\2}(z, z') = \left( \frac{C^{LL}\ti{12}}{z' - z} + \frac{C^{RL}\ti{12}}{z + z'} \right) \varphi(z')^{-1} - \left( \frac{C^{RR}\ti{12}}{z' - z} + \frac{C^{LR}\ti{12}}{z + z'} \right) \varphi(- z')^{-1},
\end{equation}
where we defined the partial split Casimirs
\begin{alignat*}{2}
C^{LL}\ti{12} &= \kappa^{ab}(T_a,0) \otimes (T_b,0), &\qquad
C^{RR}\ti{12} &= \kappa^{ab}(0,T_a) \otimes (0,T_b), \\
C^{LR}\ti{12} &= \kappa^{ab}(T_a,0) \otimes (0,T_b), &\qquad
C^{RL}\ti{12} &= \kappa^{ab}(0,T_a) \otimes (T_b,0).
\end{alignat*}
The $r/s$ form of the Poisson bracket of $\Lc(z)$ implies that the 
 Lax matrix  
 \eqref{LaxDouble} in the double Lie algebra    also has 
 a Poisson bracket of the $r/s$-form. Furthermore, it is  
 associated with the $\Rc$-matrix \eqref{RDouble} for the twist function $\varphi$ 
 given by \eqref{TwistFunction}. The projection of this Poisson bracket onto the left part of the double Lie algebra gives back the Poisson bracket for $\Lc(z)$ of the $r/s$-form discussed in section \ref{sec-3}.

\section{Analysis of the twist function and symmetries}
\label{sec-lift}

\begin{figure}[htbp]
\begin{center}
\vspace{-3pt}
\begin{tikzpicture}[scale=2]
\draw[-] (-1.2,0) to (1.2,0) ;
\draw[-] (0,-1.2) to (0,1.2) ;
\draw[solid,dashed] (0,0) circle (1) ;
\draw[solid,fill=black] (1,0) circle (0.03);
\node[right] at (1.,0.1) {$1$};
\draw[solid,fill=black] (-1,0) circle (0.03);
\node[left] at (-1,0.1) {$-1$};
\draw[-,red] (0,0) to (0.866,0.5) ;
\draw[solid,red,fill=red] (0.866,0.5) circle (0.03);
\node[red,right] at (0.9,0.5) {$z_+$};
\draw[->,blue] (0.55,0) arc (0:30:0.55);
\node[blue,right] at (0.5,0.18) { $\theta$}; 
\draw[-,red] (0,0) to (0.866,-0.5) ;
\draw[solid,red,fill=red] (0.866,-0.5) circle (0.03);
\node[red,right] at (0.9,-0.55) {$z_-$};
\draw[-,red] (0,0) to (-0.5,0.866) ;
\draw[solid,red,fill=red] (-0.5,0.866) circle (0.03);
\node[red,above] at (-0.5,0.866) { $\tilde{z}_-$};
\draw[-,red] (0,0) to (-0.5,-0.866) ;
\draw[solid,red,fill=red] (-0.5,-0.866) circle (0.03);
\node[red,below] at (-0.5,-0.866) {$\tilde{z}_+$};
\draw[->,blue] (-0.55,0) arc (0:60:-0.55);
\node[blue,left] at (-0.5,-0.3) {$\tilde{\theta}$}; 
\end{tikzpicture}
\caption{Poles of the twist function $\varphi_\bYB$ given by \eqref{TwistFunction}.\label{FigPoles}}
\end{center}
\end{figure}
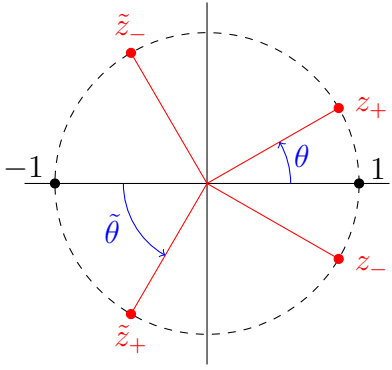

As we will see later, the poles of the twist function characterises  the 
model \cite{Vicedo:2015pna}. In the case of the bi-Yang-Baxter $\sigma$-model, the 
twist function \eqref{TwistFunction} has four simple poles, disposed on the unit 
circle of the complex plane (cf figure \ref{FigPoles}): 
\beqz 
z_\pm =  \frac{1-\frac{1}{4}(\eta^2-\etat^2) \pm i  \eta}{\zeta} =  z_\mp^* 
\et
\zt_\pm =-\frac{1+\frac{1}{4}(\eta^2-\etat^2) \pm i  \etat}{\zeta}= \zt_\mp^*.
\eeqz
Let us recall that 
\beqz
\zeta = \sqrt{\bigl(1+\qa (\eta+\etat)^2\bigr)\bigl(1+\qa(\eta-\etat)^2\bigr)}.
\eeqz
These poles can be re-expressed in a trigonometric form as 
$z_\pm=e^{\pm i\theta}$ and  $ \zt_\pm=-e^{\pm i\tilde{\theta}}$, 
with $\sin\theta = \eta/\zeta$ and $\sin\tilde{\theta} =\etat/\zeta$.


\subsection{Lax matrix at the poles of the twist function}

Evaluating the Lax matrix \eqref{Lax} at the poles of the twist function, one obtains:
\begin{subequations}\vspace{-3pt}\label{LaxPoles}
\begin{align}
\Jc_\pm & \equiv  \pm \frac{2 i K}{\eta} \Lc(z_\pm) = \pm \frac{2 i K}{\eta} \left( - j + \frac{\eta}{4K} (R_g \mp i)X \right), \\
\Jct_\pm & \equiv  \pm \frac{2 i K}{\etat} \Lc(\zt_\pm) = \pm \frac{2 i K}{\etat} \left( - \jt + \frac{\etat}{4K} (\Rt_{\gt} \mp i)X \right).
\end{align}
\end{subequations}

\noi One can verify that $\Jc_\pm$ and $\Jct_\pm$ are Poisson commuting 
Kac-Moody currents valued in $\g^\CC$ and with imaginary central charges 
\begin{align*}
\lbrace \Jc_\pm(\s)\ti{1}, \Jc_\pm(\s')\ti{2} \rbrace & =  -\left[ C\ti{12}, \Jc_\pm(\s)\ti{2} \right] \delta_{\s\s'} \pm \frac{2 i K}{\eta} C\ti{12} \delta'_{\s\s'}, \\
\lbrace \Jct_\pm(\s)\ti{1}, \Jct_\pm(\s')\ti{2} \rbrace & =  -[ C\ti{12}, \Jct_\pm(\s)\ti{2}] \delta_{\s\s'} \pm \frac{2 i K}{\etat} C\ti{12} \delta'_{\s\s'}.
\end{align*}
All the other Poisson brackets vanish. These brackets can also be seen 
more simply as a direct consequence of the $r/s$-system \eqref{SystemRS}. 
Indeed, the form \eqref{RSTwist} of the $\Rc$-matrix imposes that the values of the Lax matrix at each pole of the twist function define mutually Poisson commuting Kac-Moody currents, as already shown in~\cite{Vicedo:2015pna}.\\

Denote the gauge transformation of the Lax matrix by a $G$-valued field $h$ as 
\beqz
\Lc^h(z) = h\Lc(z)h^{-1} - h \p_\s h^{-1}.
\eeqz
One can eliminate the currents $j$ and $\jt$ in \eqref{LaxPoles} by performing 
a gauge transformation by the fields $g$ and $\gt$, respectively, 
\begin{subequations}\label{LaxPolesGauged}
\begin{align}
\Lc^g(z_\pm) & =  \frac{\eta}{4K} (R \mp i) (gXg^{-1}), \\
\Lc^{\gt}(\zt_\pm) & =  \frac{\etat}{4K} (\Rt \mp i) (\gt \Xt \gt^{-1}).
\end{align}
\end{subequations}

\subsection[Lift to the cotangent bundle $T^*L(G\times G)$]{Lift to the cotangent bundle $\bm{T^*L(G\times G)}$}
  
According to \eqref{LaxPolesGauged}, $\Lc^g(z_\pm)$ belongs to the subalgebra $\g_\mp=(R \mp i)\g$ of 
  $\g^\mathbb{C}$. Denote by $G_\mp$ the corresponding subgroup of $G^\mathbb{C}$. 
  Let $\Psi^g_\pm(\sigma)$ be a solution belonging to $G_\mp$ of 
\beqz
\partial_\sigma \Psi^g_\pm(\sigma)\, \Psi^g_\pm (\sigma)^{-1}  =  \L^g(\sigma, z^\pm).
\eeqz
Then $\Psi_\pm(\sigma) = g(\sigma)^{-1}  \Psi^g_\pm(\sigma)$ is a solution of 
\beqz
\partial_\sigma \Psi_\pm(\sigma)\, \Psi_\pm(\sigma)^{-1} =  \L(\sigma, z^\pm).
\eeqz
We recover the result that $g(\sigma)^{-1}$ 
corresponds to the first factor in 
the Iwasawa decomposition $G^\mathbb{C}=G G_\mp$ of the extended solution $\Psi_\pm(\sigma)$ 
\cite{Klimcik:2008eq,Delduc:2013fga,Vicedo:2015pna}. The same analysis 
can be carried out for $\tilde{z}_\pm$ and $\tilde{g}$. \\

Suppose we had started the construction of the 2-parameter deformation as in 
\cite{Delduc:2013fga,Delduc:2014kha,Vicedo:2015pna}. This means that we would 
have a twist function and an abstract Lax matrix, without having the expression 
of this matrix in terms of canonical fields. The analysis above proves that one 
could have derived the canonical fields $g$, $\gt$, $X$ and $\Xt$  from the values 
of the Lax matrix at the poles of the twist function.   We shall address the problem of 
constructing the corresponding 
  Hamiltonian defining the dynamics on phase space later, in subsection \ref{LiftHam}.  

\subsection{$q$-deformed symmetry algebra}

We shall 
now discuss the symmetries of the bi-Yang-Baxter $\sigma$-model. For this 
we consider the case where the fields are defined on the real line {\em i.e.} 
$\sigma \in \mathbb{R}$. Let us consider the monodromy matrices of the Lax matrix and its 
gauge transformation, at the poles $z_\pm$ of the twist function 
\begin{equation*}
T_\pm = \text{P}\overleftarrow{\text{exp}} \left( \int_{-\infty}^{+\infty} d\s \Lc(z_\pm,\s) \right), \qquad
T^g_\pm = \text{P}\overleftarrow{\text{exp}} \left( \int_{-\infty}^{+\infty} d\s \Lc^g(z_\pm,\s) \right),
\end{equation*}
and define similarly $\widetilde{T}_\pm$ and $\widetilde{T}^{\gt}_\pm$, at the poles 
$\zt_\pm$. As usual, the zero curvature equation \eqref{LaxEq} for the Lax pair implies 
the conservation of $T_\pm$ and $\widetilde{T}_\pm$. Moreover, we 
have 
\begin{equation}
T_\pm = g(+\infty)^{-1}T^g_\pm g(-\infty) \; \; \; \; \; \; \text{and} \; \; \; \; \; \; 
\widetilde{T}_\pm = \gt(+\infty)^{-1}\widetilde{T}^{\gt}_\pm \gt(-\infty).
\end{equation}
Thus, if we suppose that the boundary conditions $g(\pm\infty)$ and $\gt(\pm\infty)$ are independent of $\tau$, then $T^g_\pm$ and $\widetilde{T}^{\gt}_\pm$ are also conserved charges.

These charges are constructed as the path-ordered exponential of the currents 
$\Lc^g(z_\pm)$ and $\Lc^{\gt}(\zt_\pm)$, given by \eqref{LaxPolesGauged}. 
This  particular structure of the currents  
and the Poisson brackets \eqref{PBgX} enable one 
to show \cite{Delduc:2013fga} that 
the corresponding algebra of conserved charges forms the classical analogue of a 
quantum group.  More precisely, applying the results of~\cite{Delduc:2013fga}, one can extract from 
$T^g_\pm$ and $\widetilde{T}^{\gt}_\pm$ a set of 
  non-local charges  
which generate the Poisson algebra 
$U^{\mathcal P}_q(\g) \times U^{\mathcal P}_{\tilde{q}}(\g)$, 
analogue of a quantum group and where 
\beqz
q = \exp\left(-\frac{\eta }{4K} \right) \; \; \; \; \; \; \text{and} \; \; \; \; \; \;
\tilde{q} = \exp\left(-\frac{\etat}{4K} \right).
\eeqz
One recovers the values already indicated in \cite{Hoare:2014oua} 
 and that in the   one-paraneter 
deformation limit $\eta =\etat$  \cite{Delduc:2013fga}.

\subsection{Reconstruction of the Hamiltonian}
\label{LiftHam}

We will now show how to recover the Hamiltonian of the model from the Lax matrix and the 
twist function. Following~\cite{Vicedo:2015pna}, which treats the case of the one-parameter 
deformation $\eta=\etat$, we introduce the following Hamiltonian density\footnote{In~\cite{Vicedo:2015pna}, the expression (3.23) for $\mathcal{H}_\varphi$ contains a factor $\frac{1}{4}$. Yet this expression is for the Lax matrix in the double Lie algebra. Here, for the Lax matrix in a simple copy of $\g$, it translates to a factor $\frac{1}{2}$.}
\beqz
\mathcal{H}_\varphi(\s) = \frac{1}{2} \left( \Res_{z=0} - \Res_{z=\infty} \right) \kappa 
\left( \Lc(z,\s), \Lc(z,\s) \right) \varphi(z) \dd z.
\eeqz

\noi One can show that this Hamiltonian density can be expressed in terms of the naive Hamiltonian density \eqref{NaiveHam} and the constraint $X+\Xt$ as
\beqz
\mathcal{H}_\varphi = \mathcal{H}_0 + \kappa \bigl( \Lambda_\varphi, X+\Xt \bigr), 
\eeqz
where $\Lambda_\varphi$ is a $\g$-valued field, depending linearly 
on the fields $j$, $\jt$, $X$, $\Xt$, $R_gX$ and $\RgXt$. This Hamiltonian 
is indeed of the form \eqref{Ham}, with a fixed Lagrange multiplier 
$\Lambda_\varphi$. Thus, it gives back the correct dynamics for all the fields.

\section{Gauge fixing and Lax matrix}
\label{sec-gf}

To analyse what happens when the bi-Yang-Baxter $\sigma$-model is 
formulated as in \cite{Klimcik:2008eq}, one needs to gauge fix the $G_{\text{diag}}$ gauge invariance. 
We do this by taking $\gt =\mbox{Id}$. As already discussed in section \ref{sec-2}, 
this gauge may be reached by the field-dependent gauge transformation \eqref{GaugeSym} 
with $h=\gt$. As we shall see, this induces a gauge transformation on the Lax matrix. Let us 
first recall a general result \cite{Hou:1994hga} about the change in the Poisson 
bracket of the Lax matrix under a gauge transformation. 


\paragraph{A general result.} 
Consider  a Lax matrix taking values in $\g^\CC$ and 
whose Poisson bracket takes the $r/s$ form \eqref{RS}.
Let 
us apply a gauge transformation 
\beqz
\L\rightarrow \L^h = h \L h^{-1} - h \partial_\sigma h^{-1}
\eeqz
by some $G$-valued field $h$ constructed from the phase space fields. We suppose that the Poisson 
brackets of $h$ with itself and with the Lax matrix take the form
\begin{equation*}
\{ h_\1(\s), h_\2(\s')\}=0, \qquad
\{\L_\1(z, \s), h_\2(\s')\} h_\2(\s')^{-1} = \omega_{\1\2}(z, \s) \delta_{\s\s'},
\end{equation*}
for some $\g^\CC \otimes \g^\CC$-valued (potentially field dependent) tensor $\omega_{\1\2}(z, \s)$. 
A direct computation shows that  the Poisson bracket 
of the gauge transformed Lax matrix $\L^h(z)$ with itself is 
  also of the $r/s$-form. More precisely, one has
  \begin{subequations}\label{GTLLRh}
\begin{align}
\{\L^h_\1(z, \s),\L^h_\2(z',\s') \} &= [\R^h_{\1\2}(z, z', \s),\L^h_\1(z,\s)] \delta_{\s\s'} - [\R^h_{\2\1}(z', z, \s),\L^h_\2(z',\s)] \delta_{\s\s'}\\
&\qquad\qquad\qquad\qquad\qquad + \big( \R^h_{\1\2}(z, z', \s) + \R^h_{\2\1}(z', z, \s) \big) \delta'_{\s\s'},
\label{RSGT}
\end{align}
where the $\R$-matrix $\R^h$ is given by:
\begin{equation}
\R^h_{\1\2}(z, z', \s) = h_\1(\s) h_\2(\s)\R_{\1\2}(z, z') h_\1(\s)^{-1} h_\2(\s)^{-1} - 
h_\2(\s) \omega_{\2\1}(z',\s) h_\2(\s)^{-1}.
\end{equation}
\end{subequations}
This $\R$-matrix may be dynamical {\em i.e.} field dependent.

\paragraph{Gauge fixing as a suitable gauge transformation.}

Consider now the following gauge transformation of the Lax matrix \eqref{Lax} 
of the bi-Yang-Baxter $\sigma$-model, 
\begin{equation*}
\L^{\tilde g}(z) = \tilde g \L(z) {\tilde g}^{-1}-\tilde g\partial_\sigma{\tilde g}^{-1}.
\end{equation*}
Define then the gauge-invariant fields
\begin{equation*}
g'=g{\tilde g}^{-1},\qquad j'=g'^{-1}\partial_\sigma g'=\tilde g(j-\tilde j){\tilde g}^{-1},\qquad
X'=\tilde gX{\tilde g}^{-1},\qquad \tilde X'=\tilde g\tilde X{\tilde g}^{-1}.
\end{equation*}
Using the relation $A_{\tilde j}(z) =- A_j(z) -1$, one finds
\begin{equation*}
\L^{\tilde g}(z)=A_j(z) j' + A_X(z) X' + A_{R_g X}(z) R_{g'} X' 
+ A_{\tilde X}(z) \tilde X' + A_{\RgXt}(z)\tilde R\tilde X',
\end{equation*}
with the $A_Q$ listed in appendix \ref{CoeffJ}.  The Poisson brackets of $g'$ and $X'$ are the same as those of $g$ and $X$, but the gauge transformed constraint $X'+\tilde X'$ Poisson commute with $g'$ and $X'$. We may therefore impose the constraint $X'+\tilde X'=0$ strongly in the Lax matrix, which becomes
\begin{equation*}
\L^{\tilde g}(z) = A_j(z) j' + \big(A_X(z) - A_{\tilde X}(z) \big) X' + A_{R_g X}(z) R_{g'}X' - A_{\RgXt}(z)\tilde RX'.
\end{equation*}
The key property is that 
performing such a gauge transformation is equivalent to fixing the gauge by taking $\tilde g = \Id$ 
and replacing the canonical bracket by the Dirac bracket. Indeed, the Dirac bracket of $g$ and $X$ 
is the same as the canonical one, but the constraint $X+\tilde X$ has vanishing Dirac bracket 
with $g$ and $X$, and may thus be set strongly to zero. The gauge fixed Lax matrix is just 
$\L^{\tilde g}$. 

\paragraph{Consequence.}

Viewing the gauge fixed Lax matrix  as a suitable gauge 
transformation of the original Lax matrix   allows us to use the result \eqref{GTLLRh}. 
It leads to an easy determination of its Poisson bracket. Applying \eqref{GTLLRh} to the case at hand 
where $h = \tilde g$, we find that 
\beqz 
\omega_{\1\2}(z,\sigma)= \gt_\2(\sigma) \bigl(
A_{\Xt}(z) C_{\1\2} + A_{\RgXt}(z)  {\tilde R}_{\gt}(\sigma)_{\1\2} 
\bigr)
\gt_\2(\sigma)^{-1}.
\eeqz
As a consequence, the 
new $\R$-matrix is still non-dynamical and reads 
\begin{equation*}
\R^{\tilde g}_{\1\2}(z, z') = \frac{C_{\1\2}}{z - z'} \varphi_{\bYB}(z')^{-1} - A_{\tilde X}(z') C_{\1\2} + 
A_{\RgXt}(z')\tilde R_{\1\2}.
\end{equation*}
The new $\R$-matrix is not determined solely by the twist function and depends 
on the matrix $\Rt$ appearing in the Lagrangian. 

\section{Conclusion}

Let us end with a few comments on possible generalisations of this work.

\medskip

It was shown in \cite{Sfetsos:2015nya} that it is possible to apply a 
$\lambda$-deformation\footnote{It is also called $k$-deformation in the literature.} to the 
Yang-Baxter $\sigma$-model. Just as the $\lambda$-deformation itself is known to coincide with the 
$\sigma$-model 
obtained by combining the effects of a Poisson-Lie $T$-duality and an 
analytic continuation on the Yang-Baxter $\sigma$-model 
\cite{Vicedo:2015pna,Hoare:2015gda,Klimcik:2015gba}, 
the $\lambda$-deformation of the Yang-Baxter $\sigma$-model itself should 
also be related in a similar fashion to the bi-Yang-Baxter $\sigma$-model. 
This relation has been shown for a specific example in \cite{Sfetsos:2015nya}. 
It would be interesting to prove this in general.

\medskip

We defined in \cite{Delduc:2014uaa} a two-parameter family of integrable 
deformations of the principal chiral model on an arbitrary compact Lie group, of 
a different nature to the bi-Yang-Baxter $\sigma$-model discussed here. The two 
limits of the model defined in \cite{Delduc:2014uaa}, where one of the two parameters 
is taken to zero,  correspond to the Yang-Baxter $\sigma$-model and the principal 
chiral model with a Wess-Zumino term. As already mentioned in \cite{Hoare:2014oua}, 
one expects to be able to combine this type of deformation with a bi-Yang-Baxter 
type deformation to obtain a three-parameter deformation of the principal chiral 
model on an arbitrary Lie group. In fact, a four-parameter deformation of the 
$SU(2)$ principal chiral model has already been constructed in \cite{Lukyanov:2012zt}. Yet 
from the point of view of the twist function we only expect to be able to construct a 
three-parameter deformation in the case of an arbitrary Lie group $G$. However, recall 
that it has also been suggested in \cite{Hoare:2014pna} that the fourth parameter of the 
deformation in \cite{Lukyanov:2012zt} is related to a TsT-transformation, and therefore
shall correspond to a deformation where the twist function is not modified 
  \cite{Matsumoto:2014nra,Crichigno:2014ipa,vanTongeren:2015soa,Vicedo:2015pna}.
  
\paragraph{Acknowedgements.} 
We thank B. Hoare for very useful discussions. This work is partially 
supported by the program PICS 6412 DIGEST of CNRS and 
by the French Agence Nationale de la Recherche (ANR) under grant
 ANR-15-CE31-0006 DefIS. 

\appendix
   
\section[Coefficients $A_Q$ and $J_Q$]{Coefficients $\bm{A_Q}$ and $\bm{J_Q}$}
\label{CoeffJ}

The coefficients $A_Q(z)$ in the Lax matrix \eqref{Lax} read
\begin{subequations}\label{CoeffLax}
\begin{align}
A_j(z) &= -\frac{1}{2}\left(1+\frac{\eta^2-\etat^2}{4}+\frac{\zeta}{2}\left(z+\frac{1}{z}\right)\right) \\
A_X(z) &= -\frac{\zeta}{16K}\left(z-\frac{1}{z}\right) + f(z) \\
A_{R_gX}(z) &= \frac{\eta}{8K}\left(1+\frac{\eta^2-\etat^2}{4}+\frac{\zeta}{2}\left(z+\frac{1}{z}\right)\right) \\
A_{\jt}(z) &= -\frac{1}{2}\left(1+\frac{\etat^2-\eta^2}{4}-\frac{\zeta}{2}\left(z+\frac{1}{z}\right)\right)\\
A_{\Xt}(z) &= \frac{\zeta}{16K}\left(z-\frac{1}{z}\right) +f(z) \\
A_{\RgXt}(z) &= \frac{\etat}{8K}\left(1+\frac{\etat^2-\eta^2}{4}-\frac{\zeta}{2}\left(z+\frac{1}{z}\right)\right)
\end{align}
\end{subequations}

The coefficients $J_Q(z,z')$ in the ultralocal term \eqref{ExpectedUL} are given by
\begin{subequations} \label{ActualJd}
\begin{align}
J_j(z,z') &= -A_j(z)A_X(z')-A_j(z')A_X(z)\\
J_X(z,z') &= -A_X(z)A_X(z')+A_{R_gX}(z)A_{R_gX}(z') \\
J_{R_gX}(z,z') &= -A_X(z)A_{R_gX}(z')-A_X(z')A_{R_gX}(z) \\
J_{\jt}(z,z') &= -A_{\jt}(z)A_{\Xt}(z')-A_{\jt}(z')A_{\Xt}(z) \\
J_{\Xt}(z,z') &= -A_{\Xt}(z)A_{\Xt}(z')+A_{\RgXt}(z)A_{\RgXt}(z') \\
J_{\RgXt}(z,z') &= -A_{\Xt}(z)A_{\RgXt}(z')-A_{\Xt}(z')A_{\RgXt}(z)
\end{align}
\end{subequations}

  \providecommand{\href}[2]{#2}\begingroup\raggedright\endgroup

\end{document}